\documentclass[%
 aip,
 apl,%
 amsmath,amssymb,
 reprint,%
]{revtex4-1}

\usepackage[hidelinks]{hyperref}
\usepackage{graphicx}% Include figure files
\usepackage{dcolumn}% Align table columns on decimal point
\usepackage{bm}% bold math
\usepackage{latexsym}
\begin{document}

% \preprint{AIP/123-QED}

\title[Depletion QDs in i-Si]{Depletion-mode Quantum Dots in Intrinsic Silicon}% Force line breaks with \\
%\thanks{Footnote to title of article.}
\affiliation{ 
NanoElectronics Group, MESA+ Institute for Nanotechnology, University of Twente, P.O. Box 217,
7500 AE Enschede, The Netherlands%\\This line break forced with \textbackslash\textbackslash
}%
\author{Sergey V. Amitonov}
 \email{s.amitonov@tudelft.nl.}
\author{Paul C. Spruijtenburg}%
\author{Max W.S. Vervoort}
\author{Wilfred G. van der Wiel}
\author{Floris A. Zwanenburg}
 \email{f.a.zwanenburg@utwente.nl.}
% \homepage{http://www.Second.institution.edu/~Charlie.Author.}
%\affiliation{%
%Second institution and/or address%\\This line break forced% with \\
%}%

\date{\today}% It is always \today, today,

\begin{abstract}
We report the fabrication and electrical characterization of depletion-mode quantum dots in a two-dimensional hole gas (2DHG) in intrinsic silicon.
We use fixed charge in a SiO$_2$/Al$_2$O$_3$ dielectric stack to induce a 2DHG at the Si/SiO$_2$ interface.
Fabrication of the gate structures is accomplished with a single layer metallization process.
Transport spectroscopy reveals regular Coulomb oscillations with charging energies of $10-15$~meV
and  $3-5$~meV for the few- and many-hole regimes, respectively. 
This depletion-mode design avoids complex multilayer architectures requiring precision alignment,
and allows to adopt directly best practices already developed for depletion dots in other material systems.
%has its place alongside the enhancement-mode design 
%and could prove useful in the future experiments in intrinsic Si.
We also demonstrate a method to deactivate fixed charge in the SiO$_2$/Al$_2$O$_3$ dielectric stack using deep ultraviolet light, 
which may become an important procedure to avoid unwanted 2DHG build-up in Si~MOS quantum bits.
%
% Valid PACS numbers may be entered using the \verb+\pacs{#1}+ command.
\end{abstract}

% \pacs{Valid PACS appear here}% PACS, the Physics and Astronomy
%                              % Classification Scheme.
% \keywords{Suggested keywords}%Use showkeys class option if keyword
%                               %display desired
\maketitle

In order to perform sufficient operations in a quantum computer\cite{divincenzo_quantum_1995}, 
the quantum bits are required to be long-lived. 
Group IV semiconductors not only hold promise for very long spin coherence times
\cite{tyryshkin_electron_2011, steger_quantum_2012,veldhorst_addressable_2014, muhonen_storing_2014}, 
but also may take advantage from the scalability provided by semiconductor industry.
%feature industrial scalability applicable to quantum devices. 
These benefits have attracted much attention
\cite{liu_pauli-spin-blockade_2008,zwanenburg_spin_2009,lim_observation_2009,simmons_charge_2009,higginbotham_coherent_2014,borselli_undoped_2015} 
to quantum dots (QDs)  in group IV material systems as a framework for a solid-state scalable spin-based quantum computer\cite{loss_quantum_1998}. 
Recently hole transport in QDs became a subject of particular interest, 
both experimental\cite{li_single_2013,ares_nature_2013,spruijtenburg_single-hole_2013,li_pauli_2015,voisin_electrical_2016,watzinger_heavy-hole_2016,brauns_highly_2016}
and theoretical\cite{salfi_quantum_2016,hung_spin_2017,marcellina_spin-orbit_2017},
since the hyperfine interaction is strongly suppressed,
while the spin-orbit coupling enables all-electrical spin manipulation\cite{nowack_coherent_2007} boosting scalability of hole-based qubits.
However, enabled electrical spin control makes them vulnerable to charge noise 
that leads to dephasing and decoherence of the spin states.
Elimination of electrically active defects at the location of the QDs is essential to extend the hole spin coherence time.
In Si planar QDs, nanometer-size Coulomb islands are electrostatically
defined in a gated MOSFET-type structure at the Si/SiO$_2$ interface. 
% Despite from the quality of the crystalline Si 
The disorder and defects at this interface can be detrimental to the robustness and reliability of hole spin qubits.

\begin{figure}[]
\centering
\includegraphics[]{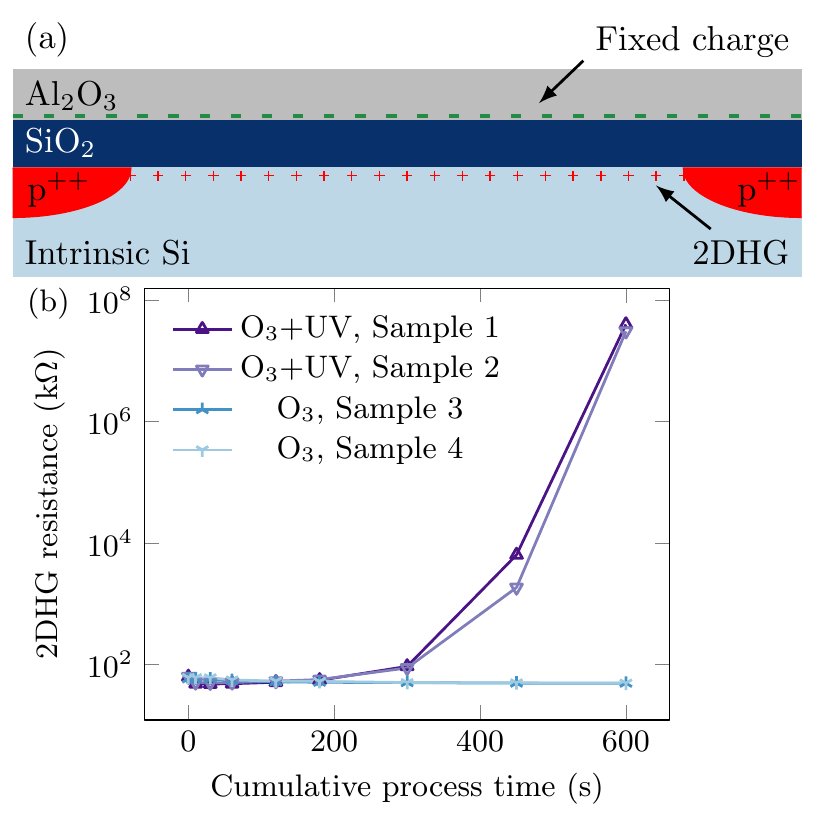}
\caption{(a)~Schematic cross-sectional view of the device structure. 
Fixed negative charge in Al$_2$O$_3$ deposited by ALD on SiO$_2$ 
induces a two-dimensional hole gas (2DHG) at the Si/SiO$_2$ interface.
The highly p-type doped regions are used to measure a resistance of the induced 2DHG.
(b)~The increase of the resistance of a 2DHG under the influence of UV light and ozone 
(``$\bigtriangleup$'',``$\bigtriangledown$''~symbols). 
Control samples exposed to ozone (``Y'',``\protect\rotatebox[origin=c]{180}{Y}''~symbols) do not show a significant change in the resistance;
measurements are done at $T$=4.2~K.}
\label{UVozone}
\end{figure}

A low-temperature ($\sim$400~$^\circ$C) hydrogen treatment is traditionally used to deactivate defects at the Si/SiO$_2$ interface. 
One way to implement it is based on hydrogen diffusion 
during atomic layer deposition (ALD) of Al$_2$O$_3$ thin films \cite{dingemans_hydrogen_2010}.
We have successfully used this approach recently\cite{spruijtenburg_passivation_2016} to improve the quality of silicon QDs.
% At the same time, it is known  that 
This method  
can lead to building up of a negative fixed charge $Q_\text{f}$ in Al$_2$O$_3$\cite{hoex_c-si_2008},
strong enough to induce 2DHG at the Si/SiO$_2$ interface.
Here we demonstrate a method to neutralize it 
% this $Q_\text{f}$  
and give a hint on the mechanism 
% shed light on the nature 
of this phenomenon.
Besides, here we utilize $Q_\text{f}$ to create depletion-mode QDs in intrinsic silicon. 

\begin{figure}[]
\centering
\includegraphics[]{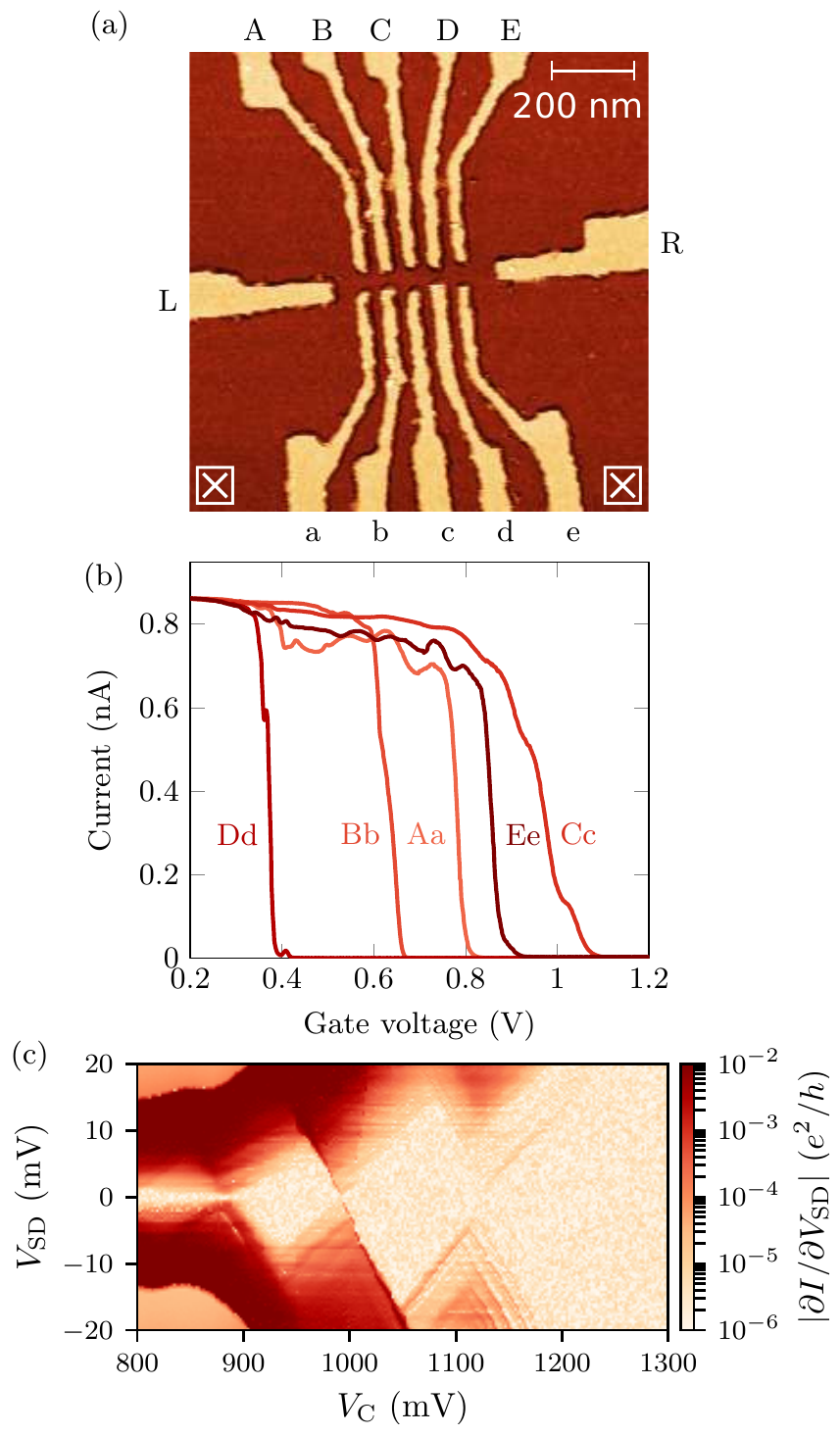}
\caption{(a)~Atomic force microscope image
of the device ``ABC'', fabricated on top of the substrate shown in Fig.~\ref{UVozone}(a). 
Source/drain ohmic contacts are depicted by~$\boxtimes$.
(b)~Current-voltage characteristics of device~``ABC'', 
individual vertical pairs of electrodes are swept together, 
all other electrodes are grounded while sweeping;
measurements are done at source-drain bias $V_\text{SD}\!=\!1$~mV and temperature $T$=4.2~K.
(c)~Numerical differential conductance $\partial I/\partial V_\text{SD}$ 
plotted vs. $V_\text{SD}$ and $V_{\text{C}}$ of a QD situated between gates B, b, C, and c. 
Only the voltage on the gate C is varied for optimal stability.
Measurements are taken at $V_\text{Aa}$=$-2$~V, $V_\text{Bb}$=0.65~V, $V_{\text{c}}$=1.52~V,
$V_\text{Dd}\!=\!V_\text{Ee}$=$0~$V, $V_\text{L}$=$-1.5$~V,  $V_\text{R}$=$-0.2$~V, and $T\approx15$~mK.
}
\label{ABC}
\end{figure}

To show the effect of negative $Q_\text{f}$ buildup, we use intrinsic Si ($\rho\geq10$~k$\Omega\cdot$cm) 
with predefined ohmic contacts to highly doped p$^{++}$ silicon areas and 7~nm of thermally grown SiO$_2$ as a substrate.
After deposition of 5~nm of Al$_2$O$_3$ by thermal ALD (TMA/H$_2$O, at $T$=250~$^\circ$C) on the substrate, we anneal it in an argon atmosphere (100~Pa, 30 minutes, $T$=400~$^\circ$C). 
Fig.~\ref{UVozone}(a) shows a schematic cross-section of the fabricated substrate. 
$Q_\text{f}$~builds up in Al$_2$O$_3$ during annealing. 
% This fixed charge induces two-dimensional hole gas (2DHG) at the Si/SiO$_2$ interface which shorts ohmic contacts.
Although there is no consensus on its origin, 
one possible explanation \cite{simon_control_2015} is that 
initial growth regime of a few first ALD cycles being non-stoichiometric 
provides an excess of oxygen atoms and deep charge traps at the SiO$_2$/Al$_2$O$_3$ interface. 
Electrons fill these traps during annealing, providing a net number of fixed charges per unit area in the
range\cite{simon_control_2015,hoex_c-si_2008} of $10^{12}-10^{13}$~cm$^{-2}$. 
This is enough to induce the two-dimensional hole gas (2DHG) at the Si/SiO$_2$ interface 
in our devices and short ohmic contacts at temperatures down to a few mK.

Besides using $Q_\text{f}$ to create depletion-mode QDs, 
we also present a method to neutralize it 
and eliminate the corresponding 2DHG using deep ultraviolet light~(UV).
We expose annealed samples in a UV ozone generator (wavelength $\lambda_\text{UV}$=254~nm)
and measure the 2-point resistance of 2DHG between ohmic contacts $R_\text{2DHG}$ \textit{ex~situ} at $T$=4.2~K.  
Fig.~\ref{UVozone}(b) shows $R_\text{2DHG}$ after 9 iterative steps of exposure 
up to a total cumulative process time of 600~s.
% after exposure to UV light. 
After exposure to UV light and O$_3$ two different samples 
(``$\bigtriangleup$'',``$\bigtriangledown$''~symbols in Fig.~\ref{UVozone}(b)) demonstrate the same behavior.
After approximately 10~minutes of cumulative exposure to UV light and O$_3$,
the $R_\text{2DHG}$ restores back to the resistance common for intrinsic Si at this temperature.
In contrast, two control samples only exposed to  O$_3$ (``Y'',``\rotatebox[origin = c]{180}{Y}''~symbols in Fig.~\ref{UVozone}(b))
demonstrate a slight decrease in $R_\text{2DHG}$.
A possible explanation of the observed neutralization is that high-energy UV radiation 
promotes diffusion of oxygen\cite{gupta_ozone_2015} in Al$_2$O$_3$, which  
% allows electrons to escape from the charge traps back to the Si substrate 
improves stoichiometry of the film, eliminating charge traps and the $Q_\text{f}$ in it, as well as the induced 2DHG.
The results in Fig.~\ref{UVozone} show that post-anneal UV exposure 
is an essential treatment of the enhancement-mode QDs in intrinsic~Si, 
as it allows to avoid short circuits via 2DHG in 
hole- and ambipolar-QDs\cite{spruijtenburg_passivation_2016,brauns_palladium_2017}.

At the same time, we propose below to take advantage of the induced 2DHG.
We see that the negative $Q_\text{f}$ present in 
Al$_2$O$_3$ after annealing induces a 2DHG without any applied gate voltage.
This 2DHG, combined with only a single layer of gate electrodes, 
is the ingredient necessary to create depletion-mode type QDs, similar to
devices created in GaAs/AlGaAs\cite{ciorga_addition_2000} or Si/SiGe heterostructures\cite{simmons_single-electron_2007,zajac_scalable_2016}. 
The following paragraphs describe a proof-of-principle of the creation of such devices in intrinsic Si, 
and feature two gate designs that can be further optimized to realize single-hole occupation.

A depletion-mode quantum dot relies on the presence of a conducting state without applying voltages to electrodes. 
The already present 2DHG is locally depleted by positive voltages on the gate electrodes to form the tunnel
barriers required for a QD \cite{ciorga_addition_2000}. 
Here we deposit metallic gates on the substrate shown in Fig.~\ref{UVozone}(a) 
using electron-beam lithography, e-beam evaporation, and lift-off techniques.
An extra layer of 5~nm Al$_2$O$_3$ is deposited on the devices using ALD to protect gates during the annealing step\cite{brauns_palladium_2017,spruijtenburg_silicon_2017}.

\begin{figure*}[]
\centering
\includegraphics[]{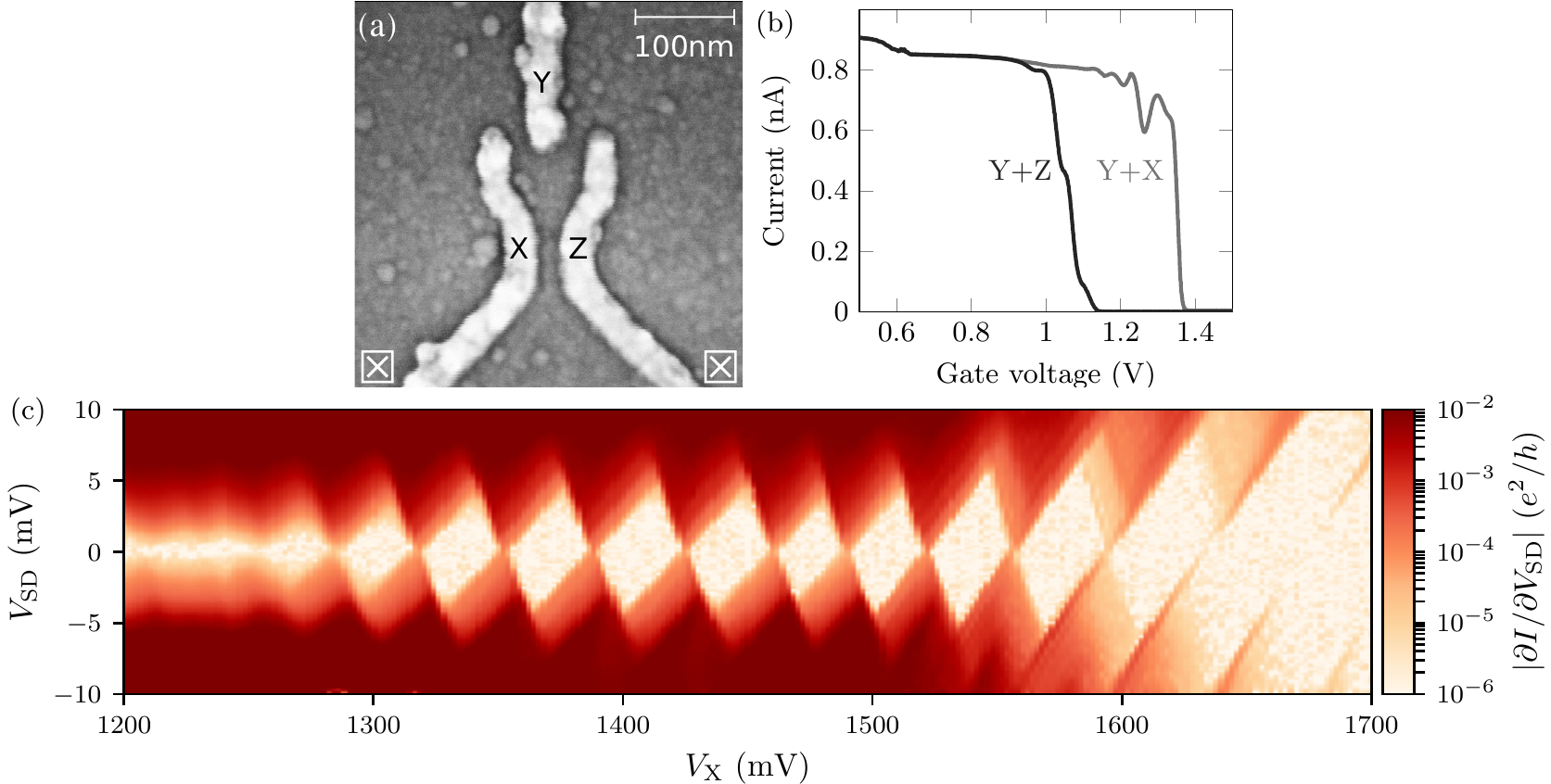}
\caption{(a)~Scanning electron microscope 
image of the device ``XYZ'', fabricated on top of the substrate shown in Fig.~\ref{UVozone}(a). 
Source/drain ohmic contacts are depicted by~$\boxtimes$.
(b)~Current-voltage characteristics of the device~``XYZ''. 
Pairs of gates X and Y or Z and Y are swept together. 
The other gate is grounded, while sweeping each pair;
measurements are performed at source-drain bias $V_\text{SD}$=1~mV and temperature $T$=4.2~K.
(c)~Numerical differential conductance $\partial I/\partial V_\text{SD}$ plotted vs. $V_\text{SD}$ and $V_\text{X}$ of QD in device ``XYZ''; 
measurements are taken at $V_\text{Y}\!=\!1425$~mV, $V_\text{Z}$=1625~mV, and $T$$\approx$300~mK.
}
\label{XYZ}
\end{figure*}

First, we test the capability of the gate electrodes to locally deplete the 2DHG and pinch off the channel.
Fig.~\ref{ABC}(a) shows the device ``ABC'' that consists of 10 electrodes 
coming in from the top and bottom.
%Electrodes opposite to each other have the same voltages applied to them,
%leading to the designation A through E for each pair, and form an individual gate. 
%If a different voltage is applied to either top or bottom electrode,
%we will append a ``t'' or ``b'' designating top or bottom, respectively.
The electrodes at the far-left (L) and far-right (R) are present in the
design for extra tunability.
Each vertical pair of the electrodes, designated the same upper and lower case letter, forms a gate. 
Fig.~\ref{ABC}(b) shows current-voltage characteristics of individual gates of the device ``ABC''.
% The rest of the electrodes during this procedure are grounded.
% Applied source-drain bias is $V_\text{SD}$=1~mV, measurements are done at $T$=4.2~K.
We can see that conduction through the channel can be turned off completely
by applying sufficient voltages on gate.
This voltage depletes the 2DHG not only underneath a pair of top and bottom electrodes but also in between them.
By tuning an applied voltage we can thus form a tunnel barrier between that pair.

Now using a couple of these tunnel barriers we will form a QD between the two adjacent gates Bb and Cc.
% The voltage on each gate is tuned to
% sufficiently high voltages that the 2DHG underneath is depleted and conduction in the channel is nearly suppressed. 
The expectation is that, due to the gate geometry, 
the gaps between B and C as well as b and c are fully depleted 
except the very center of the structure where an island is formed.
At the same time the gaps between the top and bottom electrodes remain sufficiently
conducting to act as two tunnel barriers to the formed island.
Fig.~\ref{ABC}(c) shows the transport spectroscopy of a formed QD.
%between gates Bb and Cc.
For optimal stability, only the voltage on the electrode C is varied in this measurement.
The charging energy of the last visible transition is measured to be $E_\text{C}\approx$15~meV
and is an indication of reaching the single-hole regime, 
although only a charge sensing experiment can verify single-charge occupation~\cite{elzerman_few-electron_2003,yang_dynamically_2011,yamaoka_charge_2017}.
This value is comparable to the
charging energy of previously obtained results for the last electron in Si 
\cite{spruijtenburg_passivation_2016,lim_observation_2009,yang_spin-valley_2013,lim_spin_2011}.
The results in Fig.~\ref{ABC} demonstrate that we can form QDs using this approach;
the gate design needs further optimization\cite{malkoc_optimal_2016} to improve control of a future spin qubit.

To advance results on Fig.~\ref{ABC}, we adapt a proven gate design\cite{ciorga_addition_2000} to create the device ``XYZ'', shown in Fig.~\ref{XYZ}(a).
To measure current-voltage characteristics of the left and right tunnel barriers
we ground one of the bottom gates Z or X and sweep voltage together on the two other gates
i.e. X and Y or Z and Y, respectively.
% Applied source-drain bias is $V_\text{SD}$=1~mV; measurements are done at temperature $T$=4.2~K.
% Variation in pinch-off characteristics for both devices ``ABC'' and ``XYZ''
% is explained by microscopic imperfections of the fabricated gates.
Fig.~\ref{XYZ}(c) shows transport spectroscopy of a QD formed between the gates.
Well-defined non-distorted Coulomb diamonds with charging energies of
$E_\text{C}\sim3-5$~meV demonstrate the formation of a many-hole QD.
This $E_\text{C}$ corresponds to an electrostatically defined Coulomb island 
with a size of $d$$\approx$60$-$70~nm based on a parallel plate capacitor model,
which is close to the lithographical dimensions of the QD ($\sim$75~nm).
These results show that we can form a robust hole-QD using a straightforward process and simple design.
Further reduction of the lithographical dimensions of this design, 
which is already optimized for spin-qubit performance\cite{malkoc_optimal_2016},
% down to observed in the device ``ABC'' 
% vfurther reduction of the kithographical  which is already.., could make possible reaching the  few hole regime.
may allow reaching the few-hole regime.

To summarize, we have demonstrated that we can eliminate a 2DHG induced by fixed charge in Al$_2$O$_3$ 
by a deep-UV light treatment. 
This phenomenon potentially has a great impact on the MOS electron-QD community as many groups are starting to use ALD 
and might unintentionally induce a 2DHG in their samples with no possibility to detect it.
Another important application for this technology is fine tuning of the bandgap bending 
that can improve coupling of nuclear spins to a microwave resonator\cite{pla_strain-induced_2016}. 
Finally, hybrid-mode devices, where ALD-oxide will replace source/drain accumulation gates of
the enhancement-mode devices to induce a 2DHG, simplifying fabrication and increasing the yield,
can utilize the same phenomenon.

We have also shown transport characteristics of a single-layer depletion-mode hole-QD.
Transport measurements indicate that we can fabricate QDs in the many-hole regime 
as well as QDs showing signs
of the few-hole regime, where the last visible charge-transition has a charging
energy $E_\text{C}\approx15$~meV.

Initially depletion-mode QDs were made in group IV semiconductors
 by doping, incompatible with long coherence times, or by use of epitaxial heterostructures.
The design and device, as presented here, make for a very simple system 
to create depletion-mode QDs in undoped Si without the need for accumulation gates. 
The well-defined Coulomb diamonds
when a single quantum dot is formed are promising, and optimization can push
this design even further. This optimization should take care to absorb the lessons
learned in previous designs\cite{ciorga_addition_2000}, and are an easy path to tunable arrays of QDs\cite{zajac_scalable_2016}
reaching single-hole occupation. 
Our fabrication method reliably passivates electrically active defects by hydrogen annealing,
reducing charge noise, the primary source of qubit decoherence\cite{bermeister_charge_2014}.
An additional advantage of a single-layer depletion-mode device is 
that it avoids damage from any electron-beam lithography\cite{kim_annealing_2017} in the areas where the dots are actively formed. 
The open nature of the design means that future experiments could more easily couple external
sources of light into the structure, 
for example, to optically address individual ions in silicon\cite{yin_optical_2013, morse_photonic_2017} and
transfer quantum states between distant nodes of a quantum internet\cite{kimble_quantum_2008}. 
Conversely, by using positive fixed charge in another gate dielectric stack, e.g. SiO$_2$/HfO$_2$, 
one could create electron-based depletion-mode QDs\cite{simon_control_2015}.
Given these features, the depletion-mode 
design has its place alongside the enhancement-mode design\cite{lim_observation_2009,liu_pauli-spin-blockade_2008},
and could prove extremely useful in experiments in intrinsic~Si.

\begin{acknowledgments}
We thank Matthias Brauns and Joost Ridderbos for fruitful discussions. 
We acknowledge technical support by A.A.I.~Aarnink and J.W.~Mertens.
This work is part of the research program ``Atomic physics in the solid state'' with project number 14167, 
which is (partly) financed by the Netherlands Organisation for Scientific Research (NWO).
\end{acknowledgments}

% \nocite{*}
\bibliography{Depletion_dot_paper}% Produces the bibliography via BibTeX.

\begin{thebibliography}{45}%
\makeatletter
\providecommand \@ifxundefined [1]{%
 \@ifx{#1\undefined}
}%
\providecommand \@ifnum [1]{%
 \ifnum #1\expandafter \@firstoftwo
 \else \expandafter \@secondoftwo
 \fi
}%
\providecommand \@ifx [1]{%
 \ifx #1\expandafter \@firstoftwo
 \else \expandafter \@secondoftwo
 \fi
}%
\providecommand \natexlab [1]{#1}%
\providecommand \enquote  [1]{``#1''}%
\providecommand \bibnamefont  [1]{#1}%
\providecommand \bibfnamefont [1]{#1}%
\providecommand \citenamefont [1]{#1}%
\providecommand \href@noop [0]{\@secondoftwo}%
\providecommand \href [0]{\begingroup \@sanitize@url \@href}%
\providecommand \@href[1]{\@@startlink{#1}\@@href}%
\providecommand \@@href[1]{\endgroup#1\@@endlink}%
\providecommand \@sanitize@url [0]{\catcode `\\12\catcode `\$12\catcode
  `\&12\catcode `\#12\catcode `\^12\catcode `\_12\catcode `\%12\relax}%
\providecommand \@@startlink[1]{}%
\providecommand \@@endlink[0]{}%
\providecommand \url  [0]{\begingroup\@sanitize@url \@url }%
\providecommand \@url [1]{\endgroup\@href {#1}{\urlprefix }}%
\providecommand \urlprefix  [0]{URL }%
\providecommand \Eprint [0]{\href }%
\providecommand \doibase [0]{http://dx.doi.org/}%
\providecommand \selectlanguage [0]{\@gobble}%
\providecommand \bibinfo  [0]{\@secondoftwo}%
\providecommand \bibfield  [0]{\@secondoftwo}%
\providecommand \translation [1]{[#1]}%
\providecommand \BibitemOpen [0]{}%
\providecommand \bibitemStop [0]{}%
\providecommand \bibitemNoStop [0]{.\EOS\space}%
\providecommand \EOS [0]{\spacefactor3000\relax}%
\providecommand \BibitemShut  [1]{\csname bibitem#1\endcsname}%
\let\auto@bib@innerbib\@empty
%</preamble>
\bibitem [{\citenamefont {DiVincenzo}(1995)}]{divincenzo_quantum_1995}%
  \BibitemOpen
  \bibfield  {author} {\bibinfo {author} {\bibfnamefont {D.~P.}\ \bibnamefont
  {DiVincenzo}},\ }\href
  {http://www.lanais.famaf.unc.edu.ar/cursos/quantumcomp/lectures/DiVincenzo-QC-science.pdf}
  {\bibfield  {journal} {\bibinfo  {journal} {Science}\ }\textbf {\bibinfo
  {volume} {270}},\ \bibinfo {pages} {255} (\bibinfo {year}
  {1995})}\BibitemShut {NoStop}%
\bibitem [{\citenamefont {Tyryshkin}\ \emph {et~al.}(2011)\citenamefont
  {Tyryshkin}, \citenamefont {Tojo}, \citenamefont {Morton}, \citenamefont
  {Riemann}, \citenamefont {Abrosimov}, \citenamefont {Becker}, \citenamefont
  {Pohl}, \citenamefont {Schenkel}, \citenamefont {Thewalt}, \citenamefont
  {Itoh},\ and\ \citenamefont {Lyon}}]{tyryshkin_electron_2011}%
  \BibitemOpen
  \bibfield  {author} {\bibinfo {author} {\bibfnamefont {A.~M.}\ \bibnamefont
  {Tyryshkin}}, \bibinfo {author} {\bibfnamefont {S.}~\bibnamefont {Tojo}},
  \bibinfo {author} {\bibfnamefont {J.~J.~L.}\ \bibnamefont {Morton}}, \bibinfo
  {author} {\bibfnamefont {H.}~\bibnamefont {Riemann}}, \bibinfo {author}
  {\bibfnamefont {N.~V.}\ \bibnamefont {Abrosimov}}, \bibinfo {author}
  {\bibfnamefont {P.}~\bibnamefont {Becker}}, \bibinfo {author} {\bibfnamefont
  {H.-J.}\ \bibnamefont {Pohl}}, \bibinfo {author} {\bibfnamefont
  {T.}~\bibnamefont {Schenkel}}, \bibinfo {author} {\bibfnamefont {M.~L.~W.}\
  \bibnamefont {Thewalt}}, \bibinfo {author} {\bibfnamefont {K.~M.}\
  \bibnamefont {Itoh}}, \ and\ \bibinfo {author} {\bibfnamefont {S.~A.}\
  \bibnamefont {Lyon}},\ }\href {\doibase 10.1038/nmat3182} {\bibfield
  {journal} {\bibinfo  {journal} {Nature Materials}\ }\textbf {\bibinfo
  {volume} {11}},\ \bibinfo {pages} {143} (\bibinfo {year} {2011})}\BibitemShut
  {NoStop}%
\bibitem [{\citenamefont {Steger}\ \emph {et~al.}(2012)\citenamefont {Steger},
  \citenamefont {Saeedi}, \citenamefont {Thewalt}, \citenamefont {Morton},
  \citenamefont {Riemann}, \citenamefont {Abrosimov}, \citenamefont {Becker},\
  and\ \citenamefont {Pohl}}]{steger_quantum_2012}%
  \BibitemOpen
  \bibfield  {author} {\bibinfo {author} {\bibfnamefont {M.}~\bibnamefont
  {Steger}}, \bibinfo {author} {\bibfnamefont {K.}~\bibnamefont {Saeedi}},
  \bibinfo {author} {\bibfnamefont {M.~L.~W.}\ \bibnamefont {Thewalt}},
  \bibinfo {author} {\bibfnamefont {J.~J.~L.}\ \bibnamefont {Morton}}, \bibinfo
  {author} {\bibfnamefont {H.}~\bibnamefont {Riemann}}, \bibinfo {author}
  {\bibfnamefont {N.~V.}\ \bibnamefont {Abrosimov}}, \bibinfo {author}
  {\bibfnamefont {P.}~\bibnamefont {Becker}}, \ and\ \bibinfo {author}
  {\bibfnamefont {H.-J.}\ \bibnamefont {Pohl}},\ }\href {\doibase
  10.1126/science.1217635} {\bibfield  {journal} {\bibinfo  {journal}
  {Science}\ }\textbf {\bibinfo {volume} {336}},\ \bibinfo {pages} {1280}
  (\bibinfo {year} {2012})}\BibitemShut {NoStop}%
\bibitem [{\citenamefont {Veldhorst}\ \emph {et~al.}(2014)\citenamefont
  {Veldhorst}, \citenamefont {Hwang}, \citenamefont {Yang}, \citenamefont
  {Leenstra}, \citenamefont {de~Ronde}, \citenamefont {Dehollain},
  \citenamefont {Muhonen}, \citenamefont {Hudson}, \citenamefont {Itoh},
  \citenamefont {Morello},\ and\ \citenamefont
  {Dzurak}}]{veldhorst_addressable_2014}%
  \BibitemOpen
  \bibfield  {author} {\bibinfo {author} {\bibfnamefont {M.}~\bibnamefont
  {Veldhorst}}, \bibinfo {author} {\bibfnamefont {J.~C.~C.}\ \bibnamefont
  {Hwang}}, \bibinfo {author} {\bibfnamefont {C.~H.}\ \bibnamefont {Yang}},
  \bibinfo {author} {\bibfnamefont {A.~W.}\ \bibnamefont {Leenstra}}, \bibinfo
  {author} {\bibfnamefont {B.}~\bibnamefont {de~Ronde}}, \bibinfo {author}
  {\bibfnamefont {J.~P.}\ \bibnamefont {Dehollain}}, \bibinfo {author}
  {\bibfnamefont {J.~T.}\ \bibnamefont {Muhonen}}, \bibinfo {author}
  {\bibfnamefont {F.~E.}\ \bibnamefont {Hudson}}, \bibinfo {author}
  {\bibfnamefont {K.~M.}\ \bibnamefont {Itoh}}, \bibinfo {author}
  {\bibfnamefont {A.}~\bibnamefont {Morello}}, \ and\ \bibinfo {author}
  {\bibfnamefont {A.~S.}\ \bibnamefont {Dzurak}},\ }\href {\doibase
  10.1038/nnano.2014.216} {\bibfield  {journal} {\bibinfo  {journal} {Nature
  Nanotechnology}\ }\textbf {\bibinfo {volume} {9}},\ \bibinfo {pages} {981}
  (\bibinfo {year} {2014})}\BibitemShut {NoStop}%
\bibitem [{\citenamefont {Muhonen}\ \emph {et~al.}(2014)\citenamefont
  {Muhonen}, \citenamefont {Dehollain}, \citenamefont {Laucht}, \citenamefont
  {Hudson}, \citenamefont {Kalra}, \citenamefont {Sekiguchi}, \citenamefont
  {Itoh}, \citenamefont {Jamieson}, \citenamefont {McCallum}, \citenamefont
  {Dzurak},\ and\ \citenamefont {Morello}}]{muhonen_storing_2014}%
  \BibitemOpen
  \bibfield  {author} {\bibinfo {author} {\bibfnamefont {J.~T.}\ \bibnamefont
  {Muhonen}}, \bibinfo {author} {\bibfnamefont {J.~P.}\ \bibnamefont
  {Dehollain}}, \bibinfo {author} {\bibfnamefont {A.}~\bibnamefont {Laucht}},
  \bibinfo {author} {\bibfnamefont {F.~E.}\ \bibnamefont {Hudson}}, \bibinfo
  {author} {\bibfnamefont {R.}~\bibnamefont {Kalra}}, \bibinfo {author}
  {\bibfnamefont {T.}~\bibnamefont {Sekiguchi}}, \bibinfo {author}
  {\bibfnamefont {K.~M.}\ \bibnamefont {Itoh}}, \bibinfo {author}
  {\bibfnamefont {D.~N.}\ \bibnamefont {Jamieson}}, \bibinfo {author}
  {\bibfnamefont {J.~C.}\ \bibnamefont {McCallum}}, \bibinfo {author}
  {\bibfnamefont {A.~S.}\ \bibnamefont {Dzurak}}, \ and\ \bibinfo {author}
  {\bibfnamefont {A.}~\bibnamefont {Morello}},\ }\href {\doibase
  10.1038/nnano.2014.211} {\bibfield  {journal} {\bibinfo  {journal} {Nature
  Nanotechnology}\ }\textbf {\bibinfo {volume} {9}},\ \bibinfo {pages} {986}
  (\bibinfo {year} {2014})}\BibitemShut {NoStop}%
\bibitem [{\citenamefont {Liu}\ \emph {et~al.}(2008)\citenamefont {Liu},
  \citenamefont {Fujisawa}, \citenamefont {Ono}, \citenamefont {Inokawa},
  \citenamefont {Fujiwara}, \citenamefont {Takashina},\ and\ \citenamefont
  {Hirayama}}]{liu_pauli-spin-blockade_2008}%
  \BibitemOpen
  \bibfield  {author} {\bibinfo {author} {\bibfnamefont {H.~W.}\ \bibnamefont
  {Liu}}, \bibinfo {author} {\bibfnamefont {T.}~\bibnamefont {Fujisawa}},
  \bibinfo {author} {\bibfnamefont {Y.}~\bibnamefont {Ono}}, \bibinfo {author}
  {\bibfnamefont {H.}~\bibnamefont {Inokawa}}, \bibinfo {author} {\bibfnamefont
  {A.}~\bibnamefont {Fujiwara}}, \bibinfo {author} {\bibfnamefont
  {K.}~\bibnamefont {Takashina}}, \ and\ \bibinfo {author} {\bibfnamefont
  {Y.}~\bibnamefont {Hirayama}},\ }\href {\doibase 10.1103/PhysRevB.77.073310}
  {\bibfield  {journal} {\bibinfo  {journal} {Physical Review B}\ }\textbf
  {\bibinfo {volume} {77}},\ \bibinfo {pages} {073310} (\bibinfo {year}
  {2008})}\BibitemShut {NoStop}%
\bibitem [{\citenamefont {Zwanenburg}\ \emph {et~al.}(2009)\citenamefont
  {Zwanenburg}, \citenamefont {van Rijmenam}, \citenamefont {Fang},
  \citenamefont {Lieber},\ and\ \citenamefont
  {Kouwenhoven}}]{zwanenburg_spin_2009}%
  \BibitemOpen
  \bibfield  {author} {\bibinfo {author} {\bibfnamefont {F.~A.}\ \bibnamefont
  {Zwanenburg}}, \bibinfo {author} {\bibfnamefont {C.~E. W.~M.}\ \bibnamefont
  {van Rijmenam}}, \bibinfo {author} {\bibfnamefont {Y.}~\bibnamefont {Fang}},
  \bibinfo {author} {\bibfnamefont {C.~M.}\ \bibnamefont {Lieber}}, \ and\
  \bibinfo {author} {\bibfnamefont {L.~P.}\ \bibnamefont {Kouwenhoven}},\
  }\href {\doibase 10.1021/nl803440s} {\bibfield  {journal} {\bibinfo
  {journal} {Nano Letters}\ }\textbf {\bibinfo {volume} {9}},\ \bibinfo {pages}
  {1071} (\bibinfo {year} {2009})}\BibitemShut {NoStop}%
\bibitem [{\citenamefont {Lim}\ \emph {et~al.}(2009)\citenamefont {Lim},
  \citenamefont {Zwanenburg}, \citenamefont {Huebl}, \citenamefont
  {Möttönen}, \citenamefont {Chan}, \citenamefont {Morello},\ and\
  \citenamefont {Dzurak}}]{lim_observation_2009}%
  \BibitemOpen
  \bibfield  {author} {\bibinfo {author} {\bibfnamefont {W.~H.}\ \bibnamefont
  {Lim}}, \bibinfo {author} {\bibfnamefont {F.~A.}\ \bibnamefont {Zwanenburg}},
  \bibinfo {author} {\bibfnamefont {H.}~\bibnamefont {Huebl}}, \bibinfo
  {author} {\bibfnamefont {M.}~\bibnamefont {Möttönen}}, \bibinfo {author}
  {\bibfnamefont {K.~W.}\ \bibnamefont {Chan}}, \bibinfo {author}
  {\bibfnamefont {A.}~\bibnamefont {Morello}}, \ and\ \bibinfo {author}
  {\bibfnamefont {A.~S.}\ \bibnamefont {Dzurak}},\ }\href {\doibase
  10.1063/1.3272858} {\bibfield  {journal} {\bibinfo  {journal} {Applied
  Physics Letters}\ }\textbf {\bibinfo {volume} {95}},\ \bibinfo {pages}
  {242102} (\bibinfo {year} {2009})}\BibitemShut {NoStop}%
\bibitem [{\citenamefont {Simmons}\ \emph {et~al.}(2009)\citenamefont
  {Simmons}, \citenamefont {Thalakulam}, \citenamefont {Rosemeyer},
  \citenamefont {Van~Bael}, \citenamefont {Sackmann}, \citenamefont {Savage},
  \citenamefont {Lagally}, \citenamefont {Joynt}, \citenamefont {Friesen},
  \citenamefont {Coppersmith},\ and\ \citenamefont
  {Eriksson}}]{simmons_charge_2009}%
  \BibitemOpen
  \bibfield  {author} {\bibinfo {author} {\bibfnamefont {C.~B.}\ \bibnamefont
  {Simmons}}, \bibinfo {author} {\bibfnamefont {M.}~\bibnamefont {Thalakulam}},
  \bibinfo {author} {\bibfnamefont {B.~M.}\ \bibnamefont {Rosemeyer}}, \bibinfo
  {author} {\bibfnamefont {B.~J.}\ \bibnamefont {Van~Bael}}, \bibinfo {author}
  {\bibfnamefont {E.~K.}\ \bibnamefont {Sackmann}}, \bibinfo {author}
  {\bibfnamefont {D.~E.}\ \bibnamefont {Savage}}, \bibinfo {author}
  {\bibfnamefont {M.~G.}\ \bibnamefont {Lagally}}, \bibinfo {author}
  {\bibfnamefont {R.}~\bibnamefont {Joynt}}, \bibinfo {author} {\bibfnamefont
  {M.}~\bibnamefont {Friesen}}, \bibinfo {author} {\bibfnamefont {S.~N.}\
  \bibnamefont {Coppersmith}}, \ and\ \bibinfo {author} {\bibfnamefont {M.~A.}\
  \bibnamefont {Eriksson}},\ }\href {\doibase 10.1021/nl9014974} {\bibfield
  {journal} {\bibinfo  {journal} {Nano Letters}\ }\textbf {\bibinfo {volume}
  {9}},\ \bibinfo {pages} {3234} (\bibinfo {year} {2009})}\BibitemShut
  {NoStop}%
\bibitem [{\citenamefont {Higginbotham}\ \emph {et~al.}(2014)\citenamefont
  {Higginbotham}, \citenamefont {Kuemmeth}, \citenamefont {Hanson},
  \citenamefont {Gossard},\ and\ \citenamefont
  {Marcus}}]{higginbotham_coherent_2014}%
  \BibitemOpen
  \bibfield  {author} {\bibinfo {author} {\bibfnamefont {A.~P.}\ \bibnamefont
  {Higginbotham}}, \bibinfo {author} {\bibfnamefont {F.}~\bibnamefont
  {Kuemmeth}}, \bibinfo {author} {\bibfnamefont {M.~P.}\ \bibnamefont
  {Hanson}}, \bibinfo {author} {\bibfnamefont {A.~C.}\ \bibnamefont {Gossard}},
  \ and\ \bibinfo {author} {\bibfnamefont {C.~M.}\ \bibnamefont {Marcus}},\
  }\href {\doibase 10.1103/PhysRevLett.112.026801} {\bibfield  {journal}
  {\bibinfo  {journal} {Physical Review Letters}\ }\textbf {\bibinfo {volume}
  {112}} (\bibinfo {year} {2014}),\ 10.1103/PhysRevLett.112.026801}\BibitemShut
  {NoStop}%
\bibitem [{\citenamefont {Borselli}\ \emph {et~al.}(2015)\citenamefont
  {Borselli}, \citenamefont {Eng}, \citenamefont {Ross}, \citenamefont
  {Hazard}, \citenamefont {Holabird}, \citenamefont {Huang}, \citenamefont
  {Kiselev}, \citenamefont {Deelman}, \citenamefont {Warren}, \citenamefont {{I
  Milosavljevic}}, \citenamefont {Schmitz}, \citenamefont {Sokolich},
  \citenamefont {Gyure},\ and\ \citenamefont {Hunter}}]{borselli_undoped_2015}%
  \BibitemOpen
  \bibfield  {author} {\bibinfo {author} {\bibfnamefont {M.~G.}\ \bibnamefont
  {Borselli}}, \bibinfo {author} {\bibfnamefont {K.}~\bibnamefont {Eng}},
  \bibinfo {author} {\bibfnamefont {R.~S.}\ \bibnamefont {Ross}}, \bibinfo
  {author} {\bibfnamefont {T.~M.}\ \bibnamefont {Hazard}}, \bibinfo {author}
  {\bibfnamefont {K.~S.}\ \bibnamefont {Holabird}}, \bibinfo {author}
  {\bibfnamefont {B.}~\bibnamefont {Huang}}, \bibinfo {author} {\bibfnamefont
  {A.~A.}\ \bibnamefont {Kiselev}}, \bibinfo {author} {\bibfnamefont {P.~W.}\
  \bibnamefont {Deelman}}, \bibinfo {author} {\bibfnamefont {L.~D.}\
  \bibnamefont {Warren}}, \bibinfo {author} {\bibnamefont {{I Milosavljevic}}},
  \bibinfo {author} {\bibfnamefont {A.~E.}\ \bibnamefont {Schmitz}}, \bibinfo
  {author} {\bibfnamefont {M.}~\bibnamefont {Sokolich}}, \bibinfo {author}
  {\bibfnamefont {M.~F.}\ \bibnamefont {Gyure}}, \ and\ \bibinfo {author}
  {\bibfnamefont {A.~T.}\ \bibnamefont {Hunter}},\ }\href {\doibase
  10.1088/0957-4484/26/37/375202} {\bibfield  {journal} {\bibinfo  {journal}
  {Nanotechnology}\ }\textbf {\bibinfo {volume} {26}},\ \bibinfo {pages}
  {375202} (\bibinfo {year} {2015})}\BibitemShut {NoStop}%
\bibitem [{\citenamefont {Loss}\ and\ \citenamefont
  {DiVincenzo}(1998)}]{loss_quantum_1998}%
  \BibitemOpen
  \bibfield  {author} {\bibinfo {author} {\bibfnamefont {D.}~\bibnamefont
  {Loss}}\ and\ \bibinfo {author} {\bibfnamefont {D.~P.}\ \bibnamefont
  {DiVincenzo}},\ }\href {\doibase 10.1103/PhysRevA.57.120} {\bibfield
  {journal} {\bibinfo  {journal} {Physical Review A}\ }\textbf {\bibinfo
  {volume} {57}},\ \bibinfo {pages} {120} (\bibinfo {year} {1998})}\BibitemShut
  {NoStop}%
\bibitem [{\citenamefont {Li}\ \emph {et~al.}(2013)\citenamefont {Li},
  \citenamefont {Hudson}, \citenamefont {Dzurak},\ and\ \citenamefont
  {Hamilton}}]{li_single_2013}%
  \BibitemOpen
  \bibfield  {author} {\bibinfo {author} {\bibfnamefont {R.}~\bibnamefont
  {Li}}, \bibinfo {author} {\bibfnamefont {F.~E.}\ \bibnamefont {Hudson}},
  \bibinfo {author} {\bibfnamefont {A.~S.}\ \bibnamefont {Dzurak}}, \ and\
  \bibinfo {author} {\bibfnamefont {A.~R.}\ \bibnamefont {Hamilton}},\ }\href
  {\doibase 10.1063/1.4826183} {\bibfield  {journal} {\bibinfo  {journal}
  {Applied Physics Letters}\ }\textbf {\bibinfo {volume} {103}},\ \bibinfo
  {pages} {163508} (\bibinfo {year} {2013})}\BibitemShut {NoStop}%
\bibitem [{\citenamefont {Ares}\ \emph {et~al.}(2013)\citenamefont {Ares},
  \citenamefont {Golovach}, \citenamefont {Katsaros}, \citenamefont {Stoffel},
  \citenamefont {Fournel}, \citenamefont {Glazman}, \citenamefont {Schmidt},\
  and\ \citenamefont {De~Franceschi}}]{ares_nature_2013}%
  \BibitemOpen
  \bibfield  {author} {\bibinfo {author} {\bibfnamefont {N.}~\bibnamefont
  {Ares}}, \bibinfo {author} {\bibfnamefont {V.~N.}\ \bibnamefont {Golovach}},
  \bibinfo {author} {\bibfnamefont {G.}~\bibnamefont {Katsaros}}, \bibinfo
  {author} {\bibfnamefont {M.}~\bibnamefont {Stoffel}}, \bibinfo {author}
  {\bibfnamefont {F.}~\bibnamefont {Fournel}}, \bibinfo {author} {\bibfnamefont
  {L.~I.}\ \bibnamefont {Glazman}}, \bibinfo {author} {\bibfnamefont {O.~G.}\
  \bibnamefont {Schmidt}}, \ and\ \bibinfo {author} {\bibfnamefont
  {S.}~\bibnamefont {De~Franceschi}},\ }\href {\doibase
  10.1103/PhysRevLett.110.046602} {\bibfield  {journal} {\bibinfo  {journal}
  {Physical Review Letters}\ }\textbf {\bibinfo {volume} {110}},\ \bibinfo
  {pages} {046602} (\bibinfo {year} {2013})}\BibitemShut {NoStop}%
\bibitem [{\citenamefont {Spruijtenburg}\ \emph {et~al.}(2013)\citenamefont
  {Spruijtenburg}, \citenamefont {Ridderbos}, \citenamefont {Mueller},
  \citenamefont {Leenstra}, \citenamefont {Brauns}, \citenamefont {Aarnink},
  \citenamefont {van~der Wiel},\ and\ \citenamefont
  {Zwanenburg}}]{spruijtenburg_single-hole_2013}%
  \BibitemOpen
  \bibfield  {author} {\bibinfo {author} {\bibfnamefont {P.~C.}\ \bibnamefont
  {Spruijtenburg}}, \bibinfo {author} {\bibfnamefont {J.}~\bibnamefont
  {Ridderbos}}, \bibinfo {author} {\bibfnamefont {F.}~\bibnamefont {Mueller}},
  \bibinfo {author} {\bibfnamefont {A.~W.}\ \bibnamefont {Leenstra}}, \bibinfo
  {author} {\bibfnamefont {M.}~\bibnamefont {Brauns}}, \bibinfo {author}
  {\bibfnamefont {A.~A.~I.}\ \bibnamefont {Aarnink}}, \bibinfo {author}
  {\bibfnamefont {W.~G.}\ \bibnamefont {van~der Wiel}}, \ and\ \bibinfo
  {author} {\bibfnamefont {F.~A.}\ \bibnamefont {Zwanenburg}},\ }\href
  {\doibase doi:10.1063/1.4804555} {\bibfield  {journal} {\bibinfo  {journal}
  {Applied Physics Letters}\ }\textbf {\bibinfo {volume} {102}},\ \bibinfo
  {pages} {192105} (\bibinfo {year} {2013})}\BibitemShut {NoStop}%
\bibitem [{\citenamefont {Li}\ \emph {et~al.}(2015)\citenamefont {Li},
  \citenamefont {Hudson}, \citenamefont {Dzurak},\ and\ \citenamefont
  {Hamilton}}]{li_pauli_2015}%
  \BibitemOpen
  \bibfield  {author} {\bibinfo {author} {\bibfnamefont {R.}~\bibnamefont
  {Li}}, \bibinfo {author} {\bibfnamefont {F.~E.}\ \bibnamefont {Hudson}},
  \bibinfo {author} {\bibfnamefont {A.~S.}\ \bibnamefont {Dzurak}}, \ and\
  \bibinfo {author} {\bibfnamefont {A.~R.}\ \bibnamefont {Hamilton}},\ }\href
  {\doibase 10.1021/acs.nanolett.5b02561} {\bibfield  {journal} {\bibinfo
  {journal} {Nano Letters}\ }\textbf {\bibinfo {volume} {15}},\ \bibinfo
  {pages} {7314} (\bibinfo {year} {2015})}\BibitemShut {NoStop}%
\bibitem [{\citenamefont {Voisin}\ \emph {et~al.}(2016)\citenamefont {Voisin},
  \citenamefont {Maurand}, \citenamefont {Barraud}, \citenamefont {Vinet},
  \citenamefont {Jehl}, \citenamefont {Sanquer}, \citenamefont {Renard},\ and\
  \citenamefont {De~Franceschi}}]{voisin_electrical_2016}%
  \BibitemOpen
  \bibfield  {author} {\bibinfo {author} {\bibfnamefont {B.}~\bibnamefont
  {Voisin}}, \bibinfo {author} {\bibfnamefont {R.}~\bibnamefont {Maurand}},
  \bibinfo {author} {\bibfnamefont {S.}~\bibnamefont {Barraud}}, \bibinfo
  {author} {\bibfnamefont {M.}~\bibnamefont {Vinet}}, \bibinfo {author}
  {\bibfnamefont {X.}~\bibnamefont {Jehl}}, \bibinfo {author} {\bibfnamefont
  {M.}~\bibnamefont {Sanquer}}, \bibinfo {author} {\bibfnamefont
  {J.}~\bibnamefont {Renard}}, \ and\ \bibinfo {author} {\bibfnamefont
  {S.}~\bibnamefont {De~Franceschi}},\ }\href {\doibase
  10.1021/acs.nanolett.5b02920} {\bibfield  {journal} {\bibinfo  {journal}
  {Nano Letters}\ }\textbf {\bibinfo {volume} {16}},\ \bibinfo {pages} {88}
  (\bibinfo {year} {2016})}\BibitemShut {NoStop}%
\bibitem [{\citenamefont {Watzinger}\ \emph {et~al.}(2016)\citenamefont
  {Watzinger}, \citenamefont {Kloeffel}, \citenamefont {Vukušić},
  \citenamefont {Rossell}, \citenamefont {Sessi}, \citenamefont {Kukučka},
  \citenamefont {Kirchschlager}, \citenamefont {Lausecker}, \citenamefont
  {Truhlar}, \citenamefont {Glaser}, \citenamefont {Rastelli}, \citenamefont
  {Fuhrer}, \citenamefont {Loss},\ and\ \citenamefont
  {Katsaros}}]{watzinger_heavy-hole_2016}%
  \BibitemOpen
  \bibfield  {author} {\bibinfo {author} {\bibfnamefont {H.}~\bibnamefont
  {Watzinger}}, \bibinfo {author} {\bibfnamefont {C.}~\bibnamefont {Kloeffel}},
  \bibinfo {author} {\bibfnamefont {L.}~\bibnamefont {Vukušić}}, \bibinfo
  {author} {\bibfnamefont {M.~D.}\ \bibnamefont {Rossell}}, \bibinfo {author}
  {\bibfnamefont {V.}~\bibnamefont {Sessi}}, \bibinfo {author} {\bibfnamefont
  {J.}~\bibnamefont {Kukučka}}, \bibinfo {author} {\bibfnamefont
  {R.}~\bibnamefont {Kirchschlager}}, \bibinfo {author} {\bibfnamefont
  {E.}~\bibnamefont {Lausecker}}, \bibinfo {author} {\bibfnamefont
  {A.}~\bibnamefont {Truhlar}}, \bibinfo {author} {\bibfnamefont
  {M.}~\bibnamefont {Glaser}}, \bibinfo {author} {\bibfnamefont
  {A.}~\bibnamefont {Rastelli}}, \bibinfo {author} {\bibfnamefont
  {A.}~\bibnamefont {Fuhrer}}, \bibinfo {author} {\bibfnamefont
  {D.}~\bibnamefont {Loss}}, \ and\ \bibinfo {author} {\bibfnamefont
  {G.}~\bibnamefont {Katsaros}},\ }\href {\doibase
  10.1021/acs.nanolett.6b02715} {\bibfield  {journal} {\bibinfo  {journal}
  {Nano Letters}\ }\textbf {\bibinfo {volume} {16}},\ \bibinfo {pages} {6879}
  (\bibinfo {year} {2016})}\BibitemShut {NoStop}%
\bibitem [{\citenamefont {Brauns}\ \emph {et~al.}(2016)\citenamefont {Brauns},
  \citenamefont {Ridderbos}, \citenamefont {Li}, \citenamefont {van~der Wiel},
  \citenamefont {Bakkers},\ and\ \citenamefont
  {Zwanenburg}}]{brauns_highly_2016}%
  \BibitemOpen
  \bibfield  {author} {\bibinfo {author} {\bibfnamefont {M.}~\bibnamefont
  {Brauns}}, \bibinfo {author} {\bibfnamefont {J.}~\bibnamefont {Ridderbos}},
  \bibinfo {author} {\bibfnamefont {A.}~\bibnamefont {Li}}, \bibinfo {author}
  {\bibfnamefont {W.~G.}\ \bibnamefont {van~der Wiel}}, \bibinfo {author}
  {\bibfnamefont {E.~P. A.~M.}\ \bibnamefont {Bakkers}}, \ and\ \bibinfo
  {author} {\bibfnamefont {F.~A.}\ \bibnamefont {Zwanenburg}},\ }\href
  {\doibase 10.1063/1.4963715} {\bibfield  {journal} {\bibinfo  {journal}
  {Applied Physics Letters}\ }\textbf {\bibinfo {volume} {109}},\ \bibinfo
  {pages} {143113} (\bibinfo {year} {2016})}\BibitemShut {NoStop}%
\bibitem [{\citenamefont {Salfi}\ \emph {et~al.}(2016)\citenamefont {Salfi},
  \citenamefont {Tong}, \citenamefont {Rogge},\ and\ \citenamefont
  {Culcer}}]{salfi_quantum_2016}%
  \BibitemOpen
  \bibfield  {author} {\bibinfo {author} {\bibfnamefont {J.}~\bibnamefont
  {Salfi}}, \bibinfo {author} {\bibfnamefont {M.}~\bibnamefont {Tong}},
  \bibinfo {author} {\bibfnamefont {S.}~\bibnamefont {Rogge}}, \ and\ \bibinfo
  {author} {\bibfnamefont {D.}~\bibnamefont {Culcer}},\ }\href {\doibase
  10.1088/0957-4484/27/24/244001} {\bibfield  {journal} {\bibinfo  {journal}
  {Nanotechnology}\ }\textbf {\bibinfo {volume} {27}},\ \bibinfo {pages}
  {244001} (\bibinfo {year} {2016})}\BibitemShut {NoStop}%
\bibitem [{\citenamefont {Hung}\ \emph {et~al.}(2017)\citenamefont {Hung},
  \citenamefont {Marcellina}, \citenamefont {Wang}, \citenamefont {Hamilton},\
  and\ \citenamefont {Culcer}}]{hung_spin_2017}%
  \BibitemOpen
  \bibfield  {author} {\bibinfo {author} {\bibfnamefont {J.-T.}\ \bibnamefont
  {Hung}}, \bibinfo {author} {\bibfnamefont {E.}~\bibnamefont {Marcellina}},
  \bibinfo {author} {\bibfnamefont {B.}~\bibnamefont {Wang}}, \bibinfo {author}
  {\bibfnamefont {A.~R.}\ \bibnamefont {Hamilton}}, \ and\ \bibinfo {author}
  {\bibfnamefont {D.}~\bibnamefont {Culcer}},\ }\href {\doibase
  10.1103/PhysRevB.95.195316} {\bibfield  {journal} {\bibinfo  {journal}
  {Physical Review B}\ }\textbf {\bibinfo {volume} {95}},\ \bibinfo {pages}
  {195316} (\bibinfo {year} {2017})}\BibitemShut {NoStop}%
\bibitem [{\citenamefont {Marcellina}\ \emph {et~al.}(2017)\citenamefont
  {Marcellina}, \citenamefont {Hamilton}, \citenamefont {Winkler},\ and\
  \citenamefont {Culcer}}]{marcellina_spin-orbit_2017}%
  \BibitemOpen
  \bibfield  {author} {\bibinfo {author} {\bibfnamefont {E.}~\bibnamefont
  {Marcellina}}, \bibinfo {author} {\bibfnamefont {A.~R.}\ \bibnamefont
  {Hamilton}}, \bibinfo {author} {\bibfnamefont {R.}~\bibnamefont {Winkler}}, \
  and\ \bibinfo {author} {\bibfnamefont {D.}~\bibnamefont {Culcer}},\ }\href
  {\doibase 10.1103/PhysRevB.95.075305} {\bibfield  {journal} {\bibinfo
  {journal} {Physical Review B}\ }\textbf {\bibinfo {volume} {95}} (\bibinfo
  {year} {2017}),\ 10.1103/PhysRevB.95.075305}\BibitemShut {NoStop}%
\bibitem [{\citenamefont {Nowack}\ \emph {et~al.}(2007)\citenamefont {Nowack},
  \citenamefont {Koppens}, \citenamefont {Nazarov},\ and\ \citenamefont
  {Vandersypen}}]{nowack_coherent_2007}%
  \BibitemOpen
  \bibfield  {author} {\bibinfo {author} {\bibfnamefont {K.~C.}\ \bibnamefont
  {Nowack}}, \bibinfo {author} {\bibfnamefont {F.~H.~L.}\ \bibnamefont
  {Koppens}}, \bibinfo {author} {\bibfnamefont {Y.~V.}\ \bibnamefont
  {Nazarov}}, \ and\ \bibinfo {author} {\bibfnamefont {L.~M.~K.}\ \bibnamefont
  {Vandersypen}},\ }\href {\doibase 10.1126/science.1148092} {\bibfield
  {journal} {\bibinfo  {journal} {Science}\ }\textbf {\bibinfo {volume}
  {318}},\ \bibinfo {pages} {1430} (\bibinfo {year} {2007})}\BibitemShut
  {NoStop}%
\bibitem [{\citenamefont {Dingemans}\ \emph {et~al.}(2010)\citenamefont
  {Dingemans}, \citenamefont {Beyer}, \citenamefont {van~de Sanden},\ and\
  \citenamefont {Kessels}}]{dingemans_hydrogen_2010}%
  \BibitemOpen
  \bibfield  {author} {\bibinfo {author} {\bibfnamefont {G.}~\bibnamefont
  {Dingemans}}, \bibinfo {author} {\bibfnamefont {W.}~\bibnamefont {Beyer}},
  \bibinfo {author} {\bibfnamefont {M.~C.~M.}\ \bibnamefont {van~de Sanden}}, \
  and\ \bibinfo {author} {\bibfnamefont {W.~M.~M.}\ \bibnamefont {Kessels}},\
  }\href {\doibase 10.1063/1.3497014} {\bibfield  {journal} {\bibinfo
  {journal} {Applied Physics Letters}\ }\textbf {\bibinfo {volume} {97}},\
  \bibinfo {pages} {152106} (\bibinfo {year} {2010})}\BibitemShut {NoStop}%
\bibitem [{\citenamefont {Spruijtenburg}\ \emph {et~al.}(2016)\citenamefont
  {Spruijtenburg}, \citenamefont {Amitonov}, \citenamefont {Mueller},
  \citenamefont {van~der Wiel},\ and\ \citenamefont
  {Zwanenburg}}]{spruijtenburg_passivation_2016}%
  \BibitemOpen
  \bibfield  {author} {\bibinfo {author} {\bibfnamefont {P.~C.}\ \bibnamefont
  {Spruijtenburg}}, \bibinfo {author} {\bibfnamefont {S.~V.}\ \bibnamefont
  {Amitonov}}, \bibinfo {author} {\bibfnamefont {F.}~\bibnamefont {Mueller}},
  \bibinfo {author} {\bibfnamefont {W.~G.}\ \bibnamefont {van~der Wiel}}, \
  and\ \bibinfo {author} {\bibfnamefont {F.~A.}\ \bibnamefont {Zwanenburg}},\
  }\href {\doibase 10.1038/srep38127} {\bibfield  {journal} {\bibinfo
  {journal} {Scientific Reports}\ }\textbf {\bibinfo {volume} {6}},\ \bibinfo
  {pages} {38127} (\bibinfo {year} {2016})}\BibitemShut {NoStop}%
\bibitem [{\citenamefont {Hoex}\ \emph {et~al.}(2008)\citenamefont {Hoex},
  \citenamefont {Gielis}, \citenamefont {Sanden},\ and\ \citenamefont
  {Kessels}}]{hoex_c-si_2008}%
  \BibitemOpen
  \bibfield  {author} {\bibinfo {author} {\bibfnamefont {B.}~\bibnamefont
  {Hoex}}, \bibinfo {author} {\bibfnamefont {J.~J.~H.}\ \bibnamefont {Gielis}},
  \bibinfo {author} {\bibfnamefont {M.~C. M. v.~d.}\ \bibnamefont {Sanden}}, \
  and\ \bibinfo {author} {\bibfnamefont {W.~M.~M.}\ \bibnamefont {Kessels}},\
  }\href {\doibase 10.1063/1.3021091} {\bibfield  {journal} {\bibinfo
  {journal} {Journal of Applied Physics}\ }\textbf {\bibinfo {volume} {104}},\
  \bibinfo {pages} {113703} (\bibinfo {year} {2008})}\BibitemShut {NoStop}%
\bibitem [{\citenamefont {Simon}\ \emph {et~al.}(2015)\citenamefont {Simon},
  \citenamefont {Jordan}, \citenamefont {Mikolajick},\ and\ \citenamefont
  {Dirnstorfer}}]{simon_control_2015}%
  \BibitemOpen
  \bibfield  {author} {\bibinfo {author} {\bibfnamefont {D.~K.}\ \bibnamefont
  {Simon}}, \bibinfo {author} {\bibfnamefont {P.~M.}\ \bibnamefont {Jordan}},
  \bibinfo {author} {\bibfnamefont {T.}~\bibnamefont {Mikolajick}}, \ and\
  \bibinfo {author} {\bibfnamefont {I.}~\bibnamefont {Dirnstorfer}},\ }\href
  {\doibase 10.1021/acsami.5b06606} {\bibfield  {journal} {\bibinfo  {journal}
  {ACS Applied Materials \& Interfaces}\ }\textbf {\bibinfo {volume} {7}},\
  \bibinfo {pages} {28215} (\bibinfo {year} {2015})}\BibitemShut {NoStop}%
\bibitem [{\citenamefont {Gupta}\ \emph {et~al.}(2015)\citenamefont {Gupta},
  \citenamefont {Hannah}, \citenamefont {Watson}, \citenamefont {Šutta},
  \citenamefont {Pedersen}, \citenamefont {Gadegaard},\ and\ \citenamefont
  {Gleskova}}]{gupta_ozone_2015}%
  \BibitemOpen
  \bibfield  {author} {\bibinfo {author} {\bibfnamefont {S.}~\bibnamefont
  {Gupta}}, \bibinfo {author} {\bibfnamefont {S.}~\bibnamefont {Hannah}},
  \bibinfo {author} {\bibfnamefont {C.}~\bibnamefont {Watson}}, \bibinfo
  {author} {\bibfnamefont {P.}~\bibnamefont {Šutta}}, \bibinfo {author}
  {\bibfnamefont {R.}~\bibnamefont {Pedersen}}, \bibinfo {author}
  {\bibfnamefont {N.}~\bibnamefont {Gadegaard}}, \ and\ \bibinfo {author}
  {\bibfnamefont {H.}~\bibnamefont {Gleskova}},\ }\href {\doibase
  10.1016/j.orgel.2015.03.007} {\bibfield  {journal} {\bibinfo  {journal}
  {Organic Electronics}\ }\textbf {\bibinfo {volume} {21}},\ \bibinfo {pages}
  {132} (\bibinfo {year} {2015})}\BibitemShut {NoStop}%
\bibitem [{\citenamefont {Brauns}\ \emph {et~al.}(2017)\citenamefont {Brauns},
  \citenamefont {Amitonov}, \citenamefont {Spruijtenburg},\ and\ \citenamefont
  {Zwanenburg}}]{brauns_palladium_2017}%
  \BibitemOpen
  \bibfield  {author} {\bibinfo {author} {\bibfnamefont {M.}~\bibnamefont
  {Brauns}}, \bibinfo {author} {\bibfnamefont {S.~V.}\ \bibnamefont
  {Amitonov}}, \bibinfo {author} {\bibfnamefont {P.-C.}\ \bibnamefont
  {Spruijtenburg}}, \ and\ \bibinfo {author} {\bibfnamefont {F.~A.}\
  \bibnamefont {Zwanenburg}},\ }\href {http://arxiv.org/abs/1709.07699}
  {\bibfield  {journal} {\bibinfo  {journal} {arXiv:1709.07699 [cond-mat]}\ }
  (\bibinfo {year} {2017})},\ \bibinfo {note} {arXiv: 1709.07699}\BibitemShut
  {NoStop}%
\bibitem [{\citenamefont {Ciorga}\ \emph {et~al.}(2000)\citenamefont {Ciorga},
  \citenamefont {Sachrajda}, \citenamefont {Hawrylak}, \citenamefont {Gould},
  \citenamefont {Zawadzki}, \citenamefont {Jullian}, \citenamefont {Feng},\
  and\ \citenamefont {Wasilewski}}]{ciorga_addition_2000}%
  \BibitemOpen
  \bibfield  {author} {\bibinfo {author} {\bibfnamefont {M.}~\bibnamefont
  {Ciorga}}, \bibinfo {author} {\bibfnamefont {A.~S.}\ \bibnamefont
  {Sachrajda}}, \bibinfo {author} {\bibfnamefont {P.}~\bibnamefont {Hawrylak}},
  \bibinfo {author} {\bibfnamefont {C.}~\bibnamefont {Gould}}, \bibinfo
  {author} {\bibfnamefont {P.}~\bibnamefont {Zawadzki}}, \bibinfo {author}
  {\bibfnamefont {S.}~\bibnamefont {Jullian}}, \bibinfo {author} {\bibfnamefont
  {Y.}~\bibnamefont {Feng}}, \ and\ \bibinfo {author} {\bibfnamefont
  {Z.}~\bibnamefont {Wasilewski}},\ }\href
  {http://journals.aps.org/prb/abstract/10.1103/PhysRevB.61.R16315} {\bibfield
  {journal} {\bibinfo  {journal} {Physical Review B}\ }\textbf {\bibinfo
  {volume} {61}},\ \bibinfo {pages} {R16315} (\bibinfo {year}
  {2000})}\BibitemShut {NoStop}%
\bibitem [{\citenamefont {Simmons}\ \emph {et~al.}(2007)\citenamefont
  {Simmons}, \citenamefont {Thalakulam}, \citenamefont {Shaji}, \citenamefont
  {Klein}, \citenamefont {Qin}, \citenamefont {Blick}, \citenamefont {Savage},
  \citenamefont {Lagally}, \citenamefont {Coppersmith},\ and\ \citenamefont
  {Eriksson}}]{simmons_single-electron_2007}%
  \BibitemOpen
  \bibfield  {author} {\bibinfo {author} {\bibfnamefont {C.~B.}\ \bibnamefont
  {Simmons}}, \bibinfo {author} {\bibfnamefont {M.}~\bibnamefont {Thalakulam}},
  \bibinfo {author} {\bibfnamefont {N.}~\bibnamefont {Shaji}}, \bibinfo
  {author} {\bibfnamefont {L.~J.}\ \bibnamefont {Klein}}, \bibinfo {author}
  {\bibfnamefont {H.}~\bibnamefont {Qin}}, \bibinfo {author} {\bibfnamefont
  {R.~H.}\ \bibnamefont {Blick}}, \bibinfo {author} {\bibfnamefont {D.~E.}\
  \bibnamefont {Savage}}, \bibinfo {author} {\bibfnamefont {M.~G.}\
  \bibnamefont {Lagally}}, \bibinfo {author} {\bibfnamefont {S.~N.}\
  \bibnamefont {Coppersmith}}, \ and\ \bibinfo {author} {\bibfnamefont {M.~A.}\
  \bibnamefont {Eriksson}},\ }\href {\doibase 10.1063/1.2816331} {\bibfield
  {journal} {\bibinfo  {journal} {Applied Physics Letters}\ }\textbf {\bibinfo
  {volume} {91}},\ \bibinfo {pages} {213103} (\bibinfo {year}
  {2007})}\BibitemShut {NoStop}%
\bibitem [{\citenamefont {Zajac}\ \emph {et~al.}(2016)\citenamefont {Zajac},
  \citenamefont {Hazard}, \citenamefont {Mi}, \citenamefont {Nielsen},\ and\
  \citenamefont {Petta}}]{zajac_scalable_2016}%
  \BibitemOpen
  \bibfield  {author} {\bibinfo {author} {\bibfnamefont {D.}~\bibnamefont
  {Zajac}}, \bibinfo {author} {\bibfnamefont {T.}~\bibnamefont {Hazard}},
  \bibinfo {author} {\bibfnamefont {X.}~\bibnamefont {Mi}}, \bibinfo {author}
  {\bibfnamefont {E.}~\bibnamefont {Nielsen}}, \ and\ \bibinfo {author}
  {\bibfnamefont {J.}~\bibnamefont {Petta}},\ }\href {\doibase
  10.1103/PhysRevApplied.6.054013} {\bibfield  {journal} {\bibinfo  {journal}
  {Physical Review Applied}\ }\textbf {\bibinfo {volume} {6}},\ \bibinfo
  {pages} {054013} (\bibinfo {year} {2016})}\BibitemShut {NoStop}%
\bibitem [{\citenamefont {Spruijtenburg}\ \emph {et~al.}(2017)\citenamefont
  {Spruijtenburg}, \citenamefont {Amitonov}, \citenamefont {van~der Wiel},\
  and\ \citenamefont {Zwanenburg}}]{spruijtenburg_silicon_2017}%
  \BibitemOpen
  \bibfield  {author} {\bibinfo {author} {\bibfnamefont {P.~C.}\ \bibnamefont
  {Spruijtenburg}}, \bibinfo {author} {\bibfnamefont {S.~V.}\ \bibnamefont
  {Amitonov}}, \bibinfo {author} {\bibfnamefont {W.~G.}\ \bibnamefont {van~der
  Wiel}}, \ and\ \bibinfo {author} {\bibfnamefont {F.~A.}\ \bibnamefont
  {Zwanenburg}},\ }\href {http://arxiv.org/abs/1709.08866} {\bibfield
  {journal} {\bibinfo  {journal} {arXiv:1709.08866 [cond-mat]}\ } (\bibinfo
  {year} {2017})},\ \bibinfo {note} {arXiv: 1709.08866}\BibitemShut {NoStop}%
\bibitem [{\citenamefont {Elzerman}\ \emph {et~al.}(2003)\citenamefont
  {Elzerman}, \citenamefont {Hanson}, \citenamefont {Greidanus}, \citenamefont
  {Willems~van Beveren}, \citenamefont {De~Franceschi}, \citenamefont
  {Vandersypen}, \citenamefont {Tarucha},\ and\ \citenamefont
  {Kouwenhoven}}]{elzerman_few-electron_2003}%
  \BibitemOpen
  \bibfield  {author} {\bibinfo {author} {\bibfnamefont {J.}~\bibnamefont
  {Elzerman}}, \bibinfo {author} {\bibfnamefont {R.}~\bibnamefont {Hanson}},
  \bibinfo {author} {\bibfnamefont {J.}~\bibnamefont {Greidanus}}, \bibinfo
  {author} {\bibfnamefont {L.}~\bibnamefont {Willems~van Beveren}}, \bibinfo
  {author} {\bibfnamefont {S.}~\bibnamefont {De~Franceschi}}, \bibinfo {author}
  {\bibfnamefont {L.}~\bibnamefont {Vandersypen}}, \bibinfo {author}
  {\bibfnamefont {S.}~\bibnamefont {Tarucha}}, \ and\ \bibinfo {author}
  {\bibfnamefont {L.}~\bibnamefont {Kouwenhoven}},\ }\href {\doibase
  10.1103/PhysRevB.67.161308} {\bibfield  {journal} {\bibinfo  {journal}
  {Physical Review B}\ }\textbf {\bibinfo {volume} {67}} (\bibinfo {year}
  {2003}),\ 10.1103/PhysRevB.67.161308}\BibitemShut {NoStop}%
\bibitem [{\citenamefont {Yang}\ \emph {et~al.}(2011)\citenamefont {Yang},
  \citenamefont {Lim}, \citenamefont {Zwanenburg},\ and\ \citenamefont
  {Dzurak}}]{yang_dynamically_2011}%
  \BibitemOpen
  \bibfield  {author} {\bibinfo {author} {\bibfnamefont {C.~H.}\ \bibnamefont
  {Yang}}, \bibinfo {author} {\bibfnamefont {W.~H.}\ \bibnamefont {Lim}},
  \bibinfo {author} {\bibfnamefont {F.~A.}\ \bibnamefont {Zwanenburg}}, \ and\
  \bibinfo {author} {\bibfnamefont {A.~S.}\ \bibnamefont {Dzurak}},\ }\href
  {\doibase 10.1063/1.3654496} {\bibfield  {journal} {\bibinfo  {journal} {AIP
  Advances}\ }\textbf {\bibinfo {volume} {1}},\ \bibinfo {pages} {042111}
  (\bibinfo {year} {2011})}\BibitemShut {NoStop}%
\bibitem [{\citenamefont {Yamaoka}\ \emph {et~al.}(2017)\citenamefont
  {Yamaoka}, \citenamefont {Iwasaki}, \citenamefont {Oda},\ and\ \citenamefont
  {Kodera}}]{yamaoka_charge_2017}%
  \BibitemOpen
  \bibfield  {author} {\bibinfo {author} {\bibfnamefont {Y.}~\bibnamefont
  {Yamaoka}}, \bibinfo {author} {\bibfnamefont {K.}~\bibnamefont {Iwasaki}},
  \bibinfo {author} {\bibfnamefont {S.}~\bibnamefont {Oda}}, \ and\ \bibinfo
  {author} {\bibfnamefont {T.}~\bibnamefont {Kodera}},\ }\href {\doibase
  10.7567/JJAP.56.04CK07} {\bibfield  {journal} {\bibinfo  {journal} {Japanese
  Journal of Applied Physics}\ }\textbf {\bibinfo {volume} {56}},\ \bibinfo
  {pages} {04CK07} (\bibinfo {year} {2017})}\BibitemShut {NoStop}%
\bibitem [{\citenamefont {Yang}\ \emph {et~al.}(2013)\citenamefont {Yang},
  \citenamefont {Rossi}, \citenamefont {Ruskov}, \citenamefont {Lai},
  \citenamefont {Mohiyaddin}, \citenamefont {Lee}, \citenamefont {Tahan},
  \citenamefont {Klimeck}, \citenamefont {Morello},\ and\ \citenamefont
  {Dzurak}}]{yang_spin-valley_2013}%
  \BibitemOpen
  \bibfield  {author} {\bibinfo {author} {\bibfnamefont {C.~H.}\ \bibnamefont
  {Yang}}, \bibinfo {author} {\bibfnamefont {A.}~\bibnamefont {Rossi}},
  \bibinfo {author} {\bibfnamefont {R.}~\bibnamefont {Ruskov}}, \bibinfo
  {author} {\bibfnamefont {N.~S.}\ \bibnamefont {Lai}}, \bibinfo {author}
  {\bibfnamefont {F.~A.}\ \bibnamefont {Mohiyaddin}}, \bibinfo {author}
  {\bibfnamefont {S.}~\bibnamefont {Lee}}, \bibinfo {author} {\bibfnamefont
  {C.}~\bibnamefont {Tahan}}, \bibinfo {author} {\bibfnamefont
  {G.}~\bibnamefont {Klimeck}}, \bibinfo {author} {\bibfnamefont
  {A.}~\bibnamefont {Morello}}, \ and\ \bibinfo {author} {\bibfnamefont
  {A.~S.}\ \bibnamefont {Dzurak}},\ }\href {\doibase 10.1038/ncomms3069}
  {\bibfield  {journal} {\bibinfo  {journal} {Nature Communications}\ }\textbf
  {\bibinfo {volume} {4}} (\bibinfo {year} {2013}),\
  10.1038/ncomms3069}\BibitemShut {NoStop}%
\bibitem [{\citenamefont {Lim}\ \emph {et~al.}(2011)\citenamefont {Lim},
  \citenamefont {Yang}, \citenamefont {Zwanenburg},\ and\ \citenamefont
  {Dzurak}}]{lim_spin_2011}%
  \BibitemOpen
  \bibfield  {author} {\bibinfo {author} {\bibfnamefont {W.~H.}\ \bibnamefont
  {Lim}}, \bibinfo {author} {\bibfnamefont {C.~H.}\ \bibnamefont {Yang}},
  \bibinfo {author} {\bibfnamefont {F.~A.}\ \bibnamefont {Zwanenburg}}, \ and\
  \bibinfo {author} {\bibfnamefont {A.~S.}\ \bibnamefont {Dzurak}},\ }\href
  {\doibase 10.1088/0957-4484/22/33/335704} {\bibfield  {journal} {\bibinfo
  {journal} {Nanotechnology}\ }\textbf {\bibinfo {volume} {22}},\ \bibinfo
  {pages} {335704} (\bibinfo {year} {2011})}\BibitemShut {NoStop}%
\bibitem [{\citenamefont {Malkoc}, \citenamefont {Stano},\ and\ \citenamefont
  {Loss}(2016)}]{malkoc_optimal_2016}%
  \BibitemOpen
  \bibfield  {author} {\bibinfo {author} {\bibfnamefont {O.}~\bibnamefont
  {Malkoc}}, \bibinfo {author} {\bibfnamefont {P.}~\bibnamefont {Stano}}, \
  and\ \bibinfo {author} {\bibfnamefont {D.}~\bibnamefont {Loss}},\ }\href
  {\doibase 10.1103/PhysRevB.93.235413} {\bibfield  {journal} {\bibinfo
  {journal} {Physical Review B}\ }\textbf {\bibinfo {volume} {93}} (\bibinfo
  {year} {2016}),\ 10.1103/PhysRevB.93.235413}\BibitemShut {NoStop}%
\bibitem [{\citenamefont {Pla}\ \emph {et~al.}(2016)\citenamefont {Pla},
  \citenamefont {Bienfait}, \citenamefont {Pica}, \citenamefont {Mansir},
  \citenamefont {Mohiyaddin}, \citenamefont {Morello}, \citenamefont
  {Schenkel}, \citenamefont {Lovett}, \citenamefont {Morton},\ and\
  \citenamefont {Bertet}}]{pla_strain-induced_2016}%
  \BibitemOpen
  \bibfield  {author} {\bibinfo {author} {\bibfnamefont {J.~J.}\ \bibnamefont
  {Pla}}, \bibinfo {author} {\bibfnamefont {A.}~\bibnamefont {Bienfait}},
  \bibinfo {author} {\bibfnamefont {G.}~\bibnamefont {Pica}}, \bibinfo {author}
  {\bibfnamefont {J.}~\bibnamefont {Mansir}}, \bibinfo {author} {\bibfnamefont
  {F.~A.}\ \bibnamefont {Mohiyaddin}}, \bibinfo {author} {\bibfnamefont
  {A.}~\bibnamefont {Morello}}, \bibinfo {author} {\bibfnamefont
  {T.}~\bibnamefont {Schenkel}}, \bibinfo {author} {\bibfnamefont {B.~W.}\
  \bibnamefont {Lovett}}, \bibinfo {author} {\bibfnamefont {J.~J.~L.}\
  \bibnamefont {Morton}}, \ and\ \bibinfo {author} {\bibfnamefont
  {P.}~\bibnamefont {Bertet}},\ }\href {https://arxiv.org/abs/1608.07346}
  {\bibfield  {journal} {\bibinfo  {journal} {arXiv preprint arXiv:1608.07346}\
  } (\bibinfo {year} {2016})}\BibitemShut {NoStop}%
\bibitem [{\citenamefont {Bermeister}, \citenamefont {Keith},\ and\
  \citenamefont {Culcer}(2014)}]{bermeister_charge_2014}%
  \BibitemOpen
  \bibfield  {author} {\bibinfo {author} {\bibfnamefont {A.}~\bibnamefont
  {Bermeister}}, \bibinfo {author} {\bibfnamefont {D.}~\bibnamefont {Keith}}, \
  and\ \bibinfo {author} {\bibfnamefont {D.}~\bibnamefont {Culcer}},\ }\href
  {\doibase 10.1063/1.4901162} {\bibfield  {journal} {\bibinfo  {journal}
  {Applied Physics Letters}\ }\textbf {\bibinfo {volume} {105}},\ \bibinfo
  {pages} {192102} (\bibinfo {year} {2014})}\BibitemShut {NoStop}%
\bibitem [{\citenamefont {Kim}, \citenamefont {Tyryshkin},\ and\ \citenamefont
  {Lyon}(2017)}]{kim_annealing_2017}%
  \BibitemOpen
  \bibfield  {author} {\bibinfo {author} {\bibfnamefont {J.-S.}\ \bibnamefont
  {Kim}}, \bibinfo {author} {\bibfnamefont {A.~M.}\ \bibnamefont {Tyryshkin}},
  \ and\ \bibinfo {author} {\bibfnamefont {S.~A.}\ \bibnamefont {Lyon}},\
  }\href {\doibase 10.1063/1.4979035} {\bibfield  {journal} {\bibinfo
  {journal} {Applied Physics Letters}\ }\textbf {\bibinfo {volume} {110}},\
  \bibinfo {pages} {123505} (\bibinfo {year} {2017})}\BibitemShut {NoStop}%
\bibitem [{\citenamefont {Yin}\ \emph {et~al.}(2013)\citenamefont {Yin},
  \citenamefont {Rancic}, \citenamefont {de~Boo}, \citenamefont {Stavrias},
  \citenamefont {McCallum}, \citenamefont {Sellars},\ and\ \citenamefont
  {Rogge}}]{yin_optical_2013}%
  \BibitemOpen
  \bibfield  {author} {\bibinfo {author} {\bibfnamefont {C.}~\bibnamefont
  {Yin}}, \bibinfo {author} {\bibfnamefont {M.}~\bibnamefont {Rancic}},
  \bibinfo {author} {\bibfnamefont {G.~G.}\ \bibnamefont {de~Boo}}, \bibinfo
  {author} {\bibfnamefont {N.}~\bibnamefont {Stavrias}}, \bibinfo {author}
  {\bibfnamefont {J.~C.}\ \bibnamefont {McCallum}}, \bibinfo {author}
  {\bibfnamefont {M.~J.}\ \bibnamefont {Sellars}}, \ and\ \bibinfo {author}
  {\bibfnamefont {S.}~\bibnamefont {Rogge}},\ }\href {\doibase
  10.1038/nature12081} {\bibfield  {journal} {\bibinfo  {journal} {Nature}\
  }\textbf {\bibinfo {volume} {497}},\ \bibinfo {pages} {91} (\bibinfo {year}
  {2013})}\BibitemShut {NoStop}%
\bibitem [{\citenamefont {Morse}\ \emph {et~al.}(2017)\citenamefont {Morse},
  \citenamefont {Abraham}, \citenamefont {DeAbreu}, \citenamefont {Bowness},
  \citenamefont {Richards}, \citenamefont {Riemann}, \citenamefont {Abrosimov},
  \citenamefont {Becker}, \citenamefont {Pohl}, \citenamefont {Thewalt},\ and\
  \citenamefont {Simmons}}]{morse_photonic_2017}%
  \BibitemOpen
  \bibfield  {author} {\bibinfo {author} {\bibfnamefont {K.~J.}\ \bibnamefont
  {Morse}}, \bibinfo {author} {\bibfnamefont {R.~J.~S.}\ \bibnamefont
  {Abraham}}, \bibinfo {author} {\bibfnamefont {A.}~\bibnamefont {DeAbreu}},
  \bibinfo {author} {\bibfnamefont {C.}~\bibnamefont {Bowness}}, \bibinfo
  {author} {\bibfnamefont {T.~S.}\ \bibnamefont {Richards}}, \bibinfo {author}
  {\bibfnamefont {H.}~\bibnamefont {Riemann}}, \bibinfo {author} {\bibfnamefont
  {N.~V.}\ \bibnamefont {Abrosimov}}, \bibinfo {author} {\bibfnamefont
  {P.}~\bibnamefont {Becker}}, \bibinfo {author} {\bibfnamefont {H.-J.}\
  \bibnamefont {Pohl}}, \bibinfo {author} {\bibfnamefont {M.~L.~W.}\
  \bibnamefont {Thewalt}}, \ and\ \bibinfo {author} {\bibfnamefont
  {S.}~\bibnamefont {Simmons}},\ }\href {\doibase 10.1126/sciadv.1700930}
  {\bibfield  {journal} {\bibinfo  {journal} {Science Advances}\ }\textbf
  {\bibinfo {volume} {3}},\ \bibinfo {pages} {e1700930} (\bibinfo {year}
  {2017})}\BibitemShut {NoStop}%
\bibitem [{\citenamefont {Kimble}(2008)}]{kimble_quantum_2008}%
  \BibitemOpen
  \bibfield  {author} {\bibinfo {author} {\bibfnamefont {H.~J.}\ \bibnamefont
  {Kimble}},\ }\href {\doibase 10.1038/nature07127} {\bibfield  {journal}
  {\bibinfo  {journal} {Nature}\ }\textbf {\bibinfo {volume} {453}},\ \bibinfo
  {pages} {1023} (\bibinfo {year} {2008})}\BibitemShut {NoStop}%
\end{thebibliography}%

\end{document}